\documentclass[aps,pre,10pt,superscriptaddress]{revtex4-2}
\usepackage{float}
\usepackage{epstopdf}
\usepackage{verbatim}
\usepackage{xcolor}
\usepackage{amsmath}
\usepackage{amssymb}
\usepackage[utf8]{inputenc}
\usepackage{graphicx, float,hyperref}
\usepackage{ulem}
\usepackage{caption}

\usepackage{calligra}  % For calligraphic font
\usepackage{graphicx,latexsym} 
\usepackage{ragged2e}
\AtBeginDocument{\justifying}

\begin{document}
		
		\title{{\large {\Large Role of Asymmetry in the Performance Optimization of a Relativistic Quantum Otto Engine}}}{\normalsize {\large}}
			
		\author{Monika}
		\email{monikasinghmar555@gmail.com}
			
		\author{Shishram Rebari}
		\email{rebaris@nitj.ac.in}
			
		\affiliation{Department of Physics, Dr B R Ambedkar National Institute of Technology Jalandhar, Punjab-144008, India}

\begin{abstract}
	
	We present an analytical study of the relativistic quantum Otto cycle driven by a time-dependent harmonic oscillator. By imposing an asymmetry on the two adiabatic processes of this cycle, we obtain distinct scenarios of sudden compression and sudden expansion, and analyze how asymmetry affects the performance of the relativistic quantum Otto engine. By leveraging the Omega function as a unified performance metric, we analytically characterize the efficiency in both scenarios. Our findings demonstrate that the efficiency approaches unity in the sudden compression case, while it is restricted to one-half for the sudden expansion case. Furthermore, we investigate the impact of increasing oscillator velocity on the extracted work and identify parameter regimes where either sudden compression or sudden expansion dominates. Additionally, we examine the optimal operating point using parametric efficiency-work plots, whose loop-shaped structure shows that increasing oscillator velocity enhances both work output and efficiency. Finally, through a detailed phase diagram analysis of the Otto cycle, we observe that the operational region corresponding to the engine mode expands with increasing oscillator velocity, while the refrigeration regime shrinks correspondingly.
	
\end{abstract}

	\maketitle

\section{Introduction}
	 
Thermal devices play a crucial role in modern technological and industrial developments, forming the backbone of energy conversion and thermal management. Heat engines are key thermal devices for converting heat from a hot thermal reservoir into mechanical work, and their performance is limited by the Carnot efficiency, given by $\eta_{c} = 1 - T_c/T_h $, where $T_c$ and $T_h$ are the temperatures of the cold and hot reservoirs, respectively \cite{	binder2018thermodynamics, myers2022quantum, cengel2002thermodynamics, borgnakke2025fundamentals}. However, Carnot efficiency is achieved only in the quasistatic limit, where processes are reversible and infinitely slow, thereby limiting its applicability to realistic, finite-time operations. However, the rapid advancement of quantum technologies has extended thermodynamic concepts into the quantum regime \cite{singh2020multi, kaur2025performance, das2020quantum}. In this regime, quantum effects such as entanglement, coherence, and quantum correlations can significantly affect the performance of thermal devices \cite{brunner2014entanglement, camati2019coherence, shi2020quantum}. Therefore, quantum heat engines have attracted considerable attention to investigate energy conversion at the microscopic scale.

Most studies on quantum heat engines have been conducted within non-relativistic frameworks. Current research has drawn attention to relativistic quantum heat engines, which are largely focused on the feasibility of work extraction and the development of operational protocols \cite{Relativistic2025, Papadatos2021, myers2021quantum}. The Unruh effect establishes a strong connection between relativistic motion and thermalization,  wherein uniform acceleration gives rise to a thermal response equivalent to that of an effective heat reservoir \cite{aries2018, gray2018, xu2020, kane2021}. More recently, it has been shown that relativistic motion itself can serve as a quantum resource, enabling efficiencies beyond the classical Carnot limit \cite{Relativistic2025, Moustos2025}. 

Recently, the quantum Otto cycle with a time-dependent harmonic oscillator as the working material has been extensively studied because of its conceptual simplicity and experimental relevance \cite{kosloff2017, saryal2021, PhysRevE.106.024137, kaneyasu2023, shastri2022}. Furthermore, most analyses are restricted to either adiabatic or non-adiabatic driving. In these symmetric configurations, the cycle yields negligible power output when both work strokes are adiabatic, whereas it exhibits reduced efficiency due to quantum friction when both are non-adiabatic. To overcome these competing limitations, an extension of the standard quantum Otto cycle is required that incorporates asymmetric configurations, in which the two work strokes are treated differently: one is sudden, and the other is adiabatic \cite{singh2024asymmetric}. A comprehensive understanding of the role of such asymmetry on the performance of the relativistic quantum cycle remains lacking. Further, in finite-time operations, the engine efficiency alone is not always the most appropriate performance measure, as maximizing it often results in vanishing power output, which is unfavorable for practical applications. This trade-off between efficiency and power has motivated the introduction of optimization criteria \cite{kaur2024optimization, monika2024quantum, monika2024ecological, singh2020optimal, zhang2020optimal}. One such criterion is the Omega ($\Omega$) function, which provides a unified framework to balance the maximum useful energy and the minimum lost energy, making it a valuable tool for performance optimization in quantum thermal devices \cite{hernandez2001unified, singh2022unified, kaur2021unified, zhang2020}. In this work, we address these shortcomings by doing a comprehensive analytical study of the asymmetric relativistic quantum Otto cycle with a time-dependent harmonic oscillator as the working medium. We consider two different asymmetric scenarios: sudden compression and sudden expansion, and derive analytical expressions for the efficiency at the optimal value of the $\Omega$ function for both cases in the high-temperature limit. By constructing the complete phase diagram as a function of system parameters, we identify the regions corresponding to the different operational modes: engine, refrigerator, heater, and thermal accelerator.

The structure of the present paper is as follows:
Section \ref{sa} presents the theoretical model of the relativistic quantum Otto cycle, employing a harmonic oscillator as the working substance. Section \ref{sb} is devoted to the analytical derivation of the efficiency at the maximum $\Omega$ function. In subsection \ref{sb1}, we investigate the engine performance during the sudden compression stroke, while subsection \ref{sb2} presents the corresponding analysis for the sudden expansion stroke. Section \ref{sc} addresses the complete phase diagram of the Otto cycle for various values of oscillator velocity. Finally, Section \ref{sd} summarizes and concludes the findings.
	
	\section{Quantum Otto Cycle} \label{sa}
	
	\begin{figure}
		\centering
		\includegraphics[width=8.8cm]{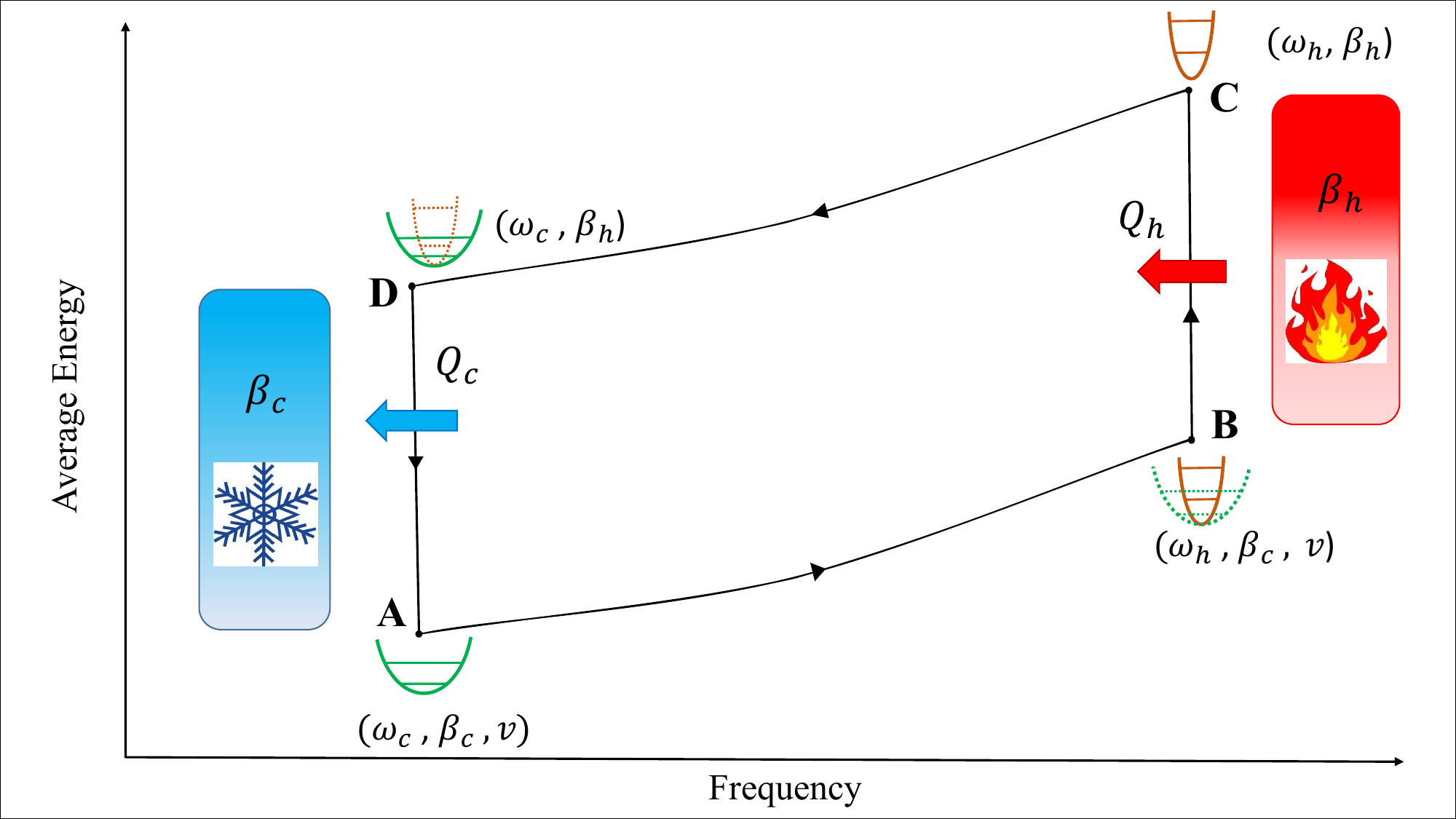}
		\caption{Diagrammatic illustration of a quantum Otto cycle based on a time-dependent harmonic oscillator.}
		\label{fig.1}
	\end{figure}
	
   We consider a relativistic quantum Otto cycle with an Unruh–DeWitt detector, modeled as a harmonic oscillator and serving as the working medium. The cycle has four consecutive processes, which are briefly discussed as follows (see Fig. \ref{fig.1}) \cite{Shaghaghi2025, PhysRevE.106.024137}: 
 
 1). Adiabatic Compression (A $\to$ B): Initially, the harmonic oscillator is in a thermal state at inverse temperature $\beta_{c}$ and moves with constant velocity $v$. The system is isolated from the reservoirs, and the oscillator frequency is changed from $\omega_{c}$ to $\omega_{h}$ via an external agency. Work is performed on the system, and no heat exchange takes place during this process.
 
 2). Hot Isochore (B $\to$ C): Now, the oscillator is at fixed frequency $\omega_{h}$ and is subjected to uniform acceleration. Due to the Unruh effect, the detector perceives an effective thermal reservoir at inverse temperature $\beta_{h}$ = $(k_B T_{h})^{-1}$. Heat flows into the system, and no work is performed during this process.
 
 3). Adiabatic Expansion (C $\to$ D): During this process, the system is again isolated, and no heat exchange takes place. The oscillator frequency returns from $\omega_{h}$ to its initial value $\omega_{c}$, and the work is extracted.
 
 4). Cold Isochore (D $\to$ A): In the final process, the oscillator moves with constant velocity $v$ at fixed frequency $\omega_{c}$. In the absence of acceleration, the detector experiences an effective cold reservoir at inverse temperature $\beta_{c}$ = $(k_B T_{c})^{-1}$. Heat flows out of the system, and the system relaxes back to its initial thermal state.
	
During these four processes of the cycle, the average energies of the oscillator, $\langle H \rangle_{i}$ ($i = A, B, C, D$), are given by ($\ k_{B}$ = $\hbar = 1$) \cite{Shaghaghi2025,singh2024asymmetric}:
	
\begin{align}
	\langle H \rangle_A &= \dfrac{\sqrt{1-v^2}}{2\beta_c v}
	\ln \left[
	\dfrac{\sinh \left(\dfrac{\beta_c \omega_c}{2}\sqrt{\dfrac{1+v}{1-v}}\right)}
	{\sinh \left(\dfrac{\beta_c \omega_c}{2}\sqrt{\dfrac{1-v}{1+v}}\right)}
	\right], \label{eq:1} \\
	\langle H \rangle_B &= \dfrac{\omega_{h}\sqrt{1-v^2}}{2\omega_{c}\beta_c v} \lambda_{AB}
	\ln \left[\dfrac{\sinh \left(\frac{\beta_c \omega_c}{2}\sqrt{\dfrac{1+v}{1-v}}\right)}
	{\sinh \left(\dfrac{\beta_c \omega_c}{2}\sqrt{\dfrac{1-v}{1+v}}\right)}
	\right], \label{eq:2} \\
	\langle H_C \rangle &= \dfrac{\omega_h}{2} \coth \left(\dfrac{\beta_h \omega_h}{2}\right), \label{eq:3} \\
	\langle H_D \rangle &= \frac{\omega_c}{2} \lambda_{CD} \coth \left(\frac{\beta_h \omega_h}{2}\right), \label{eq:4} 
\end{align}
where $\lambda_{i}$ ($i = AB, CD$) is the adiabaticity parameter and in general, $\lambda_i \geq 1$. For adiabatic cases, $\lambda_i$ = 1, whereas for sudden-switch cases, $\lambda_i$ = ($\omega^2_{c}+\omega^2_{h})/2 \omega_{c} \omega_{h} > 1$. \\
The heat exchanged during the hot and cold isochoric processes is given, respectively, by:
\begin{align}
	\langle Q \rangle_{h} &= \langle H \rangle_C - \langle H \rangle_B \nonumber \\
	&= \dfrac{\omega_h}{2} \left(\coth\left(\dfrac{\beta_h \omega_h}{2}\right)- \dfrac{\sqrt{1-v^2}}{\beta_c \omega_c v} \lambda_{AB} \ln \left[\dfrac{\sinh\left(\dfrac{\beta_c \omega_c}{2}\sqrt{\dfrac{1+v}{1-v}}\right)}{\sinh\left(\dfrac{\beta_c \omega_c}{2}\sqrt{\dfrac{1-v}{1+v}}\right)}\right]\right), \label{eq:5}
\end{align}
\begin{align}
	\langle Q \rangle_{c} &= \langle H \rangle_{A} - \langle H \rangle_{D} \nonumber \\
	&= \dfrac{\sqrt{1-v^2}}{2\beta_c v}
	\ln \left[
	\dfrac{\sinh \left(\dfrac{\beta_c \omega_c}{2}\sqrt{\dfrac{1+v}{1-v}}\right)}
	{\sinh \left(\dfrac{\beta_c \omega_c}{2}\sqrt{\dfrac{1-v}{1+v}}\right)}
	\right]-\frac{\omega_{c}}{2}\lambda_{CD} \coth \left(\frac{\beta_{h} \omega_{h}}{2}\right). \label{eq:6}
\end{align}
We employ a sign convention in which heat absorbed (or released) by the system is positive (or negative). Under this convention, the first law of thermodynamics yields the net work extracted from the engine as $ \langle W \rangle_{ext} = \langle Q \rangle_c + \langle Q \rangle_h $.

	\section{Asymmetric Relativistic Quantum Otto Engine} \label{sb}
	
	The heat engine efficiency is defined as the fraction of net extracted work to the absorbed heat from the hot reservoir and is given by
	\begin{equation}
		\eta = \dfrac{ \langle W \rangle_{ext}}{\langle Q \rangle_h}. \label{eq:7}
	\end{equation}
	This section is devoted to the optimization of the asymmetric relativistic Otto engine within the framework of the $\Omega$ function, which balances the maximum useful energy and the minimum lost energy and is expressed as \cite{PhysRevE.106.024137, monika2025asymmetric, monika2026optimal}:  
	\begin{equation}
		\Omega = 2W_{ext}-\eta_{max}Q_{h},\label{eq:8}
	\end{equation}
	where $\eta_{max}$ denotes the maximum attainable efficiency of the engine and satisfies $\eta_{\max} \leq \eta_{c}$, where $\eta_{c}$ is the Carnot efficiency. By introducing asymmetry between the adiabatic processes of the Otto cycle, we identify two distinct operational regimes: sudden compression and sudden expansion \cite{monika2025asymmetric}. We develop the analysis for the sudden compression stroke first, and then systematically extend our findings to the sudden expansion case.

	\subsection{Sudden compression stroke} \label{sb1}
	
	In the present scenario, we consider the compression stroke ($A$ $\to$ $B$) to be sudden, while the expansion stroke ($C$ $\to$ $D$) is treated as adiabatic. The adiabaticity parameters associated with these strokes take the values $\lambda_{AB}$ = $(\omega^2_{c}+\omega^2_{h})/2 \omega_{c} \omega_{h}$ and $\lambda_{CD}$ = 1, respectively \cite{singh2024asymmetric}. To derive analytical expressions, we examine the model in the high-temperature regime for which $\coth \left(\beta_{i}\omega_{i}/2\right) \approx  2/\beta_{i}\omega_{i} $ ($i = c, h$) and $ \sinh x \approx x$  (here, $x = \beta_{c} \omega_{c} \sqrt{(1+v)/(1-v)}/2) $. Within this approximation, we can express the absorbed heat and the extracted work, respectively, as:
	\begin{equation}
		Q_{h}^{SC} = \dfrac{1}{2 z^2\beta_{h}}\left[2z^2 - \tau (z^2+1) f_v\right]\label{eq:9},
	\end{equation}
	\begin{equation}
		W_{SC}^{HT} = \dfrac{(1-z)}{2 z^2 \beta_{h}}\left(2 z^2-\tau (1+z) f_v\right),\label{eq:10}
	\end{equation}
  	where $z =\omega_{c}/\omega_{h}$, $\tau$ = $\beta_{h}$/$\beta_{c}$, and $f_v = \sqrt{1-v^2} \ln \left[(1+v)/(1-v)\right]/(2v) $. By using Eq. (\ref{eq:7}), the efficiency expression for this scenario is \cite{singh2024asymmetric}:
	\begin{equation}
		\eta^{HT}_{SC} = \dfrac{(1-z)\left[2z^{2}-\tau (1+z)f_v\right]}{2z^{2}-\tau (z^2+1)f_v}.\label{eq:11}
	\end{equation}
	To obtain the maximum efficiency, we optimize Eq. (\ref{eq:11}) with respect to $z$, i.e., by solving $\partial$$\eta_{SC}^{HT}$/$\partial z = 0$, resulting in the following cubic equation:
	\begin{equation}
		z^3 (\tau v x_a \sqrt{1-v^2} - 4v^2)+3 \tau v x_a \sqrt{1-v^2} z + (v^2-1)\tau^2 x_a^2  = 0, \label{eq:12}
	\end{equation}
	where $x_a = \ln \left[(1+v)/(1-v)\right]$. We cannot solve the above equation in terms of real radicals. Therefore, we employ the method of casus irreducibilis to solve the cubic equation and obtain its solution in terms of inverse trigonometric functions as follows (see \ref{appendixA}) \cite{benenti2017fundamental, singh2024asymmetric}:
	\begin{equation}
		z^*_{SC} = 2\sqrt{\dfrac{\tau x_a \sqrt{1-v^2}}{4v-\tau x_a \sqrt{1-v^2}}}\cos \left[\dfrac{1}{3}\cos^{-1}\left(\dfrac{\tau^2 x_a^2(v^2-1)\sqrt{4v^2-\tau v x_a \sqrt{1-v^2}}}{2(\tau v x_a\sqrt{1-v^2})^{3/2}}\right)\right].\label{eq:13}
	\end{equation}
	Upon substituting this optimal value of $z$ (Eq. (\ref{eq:13})) into Eq. (\ref{eq:11}), the maximum efficiency ($\eta_{SC}^{max}$) is found to be:
	\begin{equation}
		\eta^{max}_{SC} = \dfrac{(F-1)\left[(1-\eta_{c})(F+1)f_v-2F^2\right]}{2F^2-(1-\eta_{c})(F^2+1)f_v} ,\label{eq:14}
	\end{equation}
	where $F = 2 \sqrt{\frac{E}{4v - E}} \cos\left[\frac{1}{3} \cos^{-1} \left(
	\frac{E^{2}\sqrt{4v^{2}-vE}}{2 (vE)^{3/2}}\right)\right]$ and $E = (1-\eta_{c}) x_a \sqrt{1-v^2}$. Now, substituting Eqs. (\ref{eq:9}), (\ref{eq:10}), and (\ref{eq:14}) into Eq. (\ref{eq:8}) yields the $\Omega$ function, whose optimization with respect to $z$ gives:
	
	\begin{equation}
		\small
		z^{*(\Omega)}_{SC} = \dfrac{ \left[\tau x_a \left(
			\begin{aligned}
				& 2\cos\left(\dfrac{2}{3}\cos^{-1}G\right)
				\left[\tau x_c (1-v^2)\left(16v-3\tau x_c \sqrt{1-v^2}\right)-16 v^2 \sqrt{1-v^2}\right] \\
				& -24 v G \cos \left(\dfrac{1}{3}\cos^{-1}G\right)
				\left[4v\sqrt{1-v^2}+\tau x_a (v^2-1)\right] \\
				& -16v^2\sqrt{1-v^2}
				+\tau x_c (1-v^2)(40v-9\tau x_c\sqrt{1-v^2})
			\end{aligned}
			\right)\right]^{1/3}
		}
		{\biggl\{4v\left[1+2 \cos \left(\dfrac{2}{3}\cos^{-1}G\right)\right] \left[\tau x_a\left(\tau x_a (v^2-1)+8v\sqrt{1-v^2}\right)-16v^2\right]\biggr\}^{1/3}},
		\label{eq:15}
	\end{equation}
	where $G = -x_b/\left[2v\sqrt{x_b/(4v-x_b)}\right]$, $x_b = \tau x_a \sqrt{1-v^2}$, and $x_c = \ln\left[1/(1+v)\right]+\ln(1+v) $. 
	\begin{figure}
		\centering
		\includegraphics[width=8.8cm]{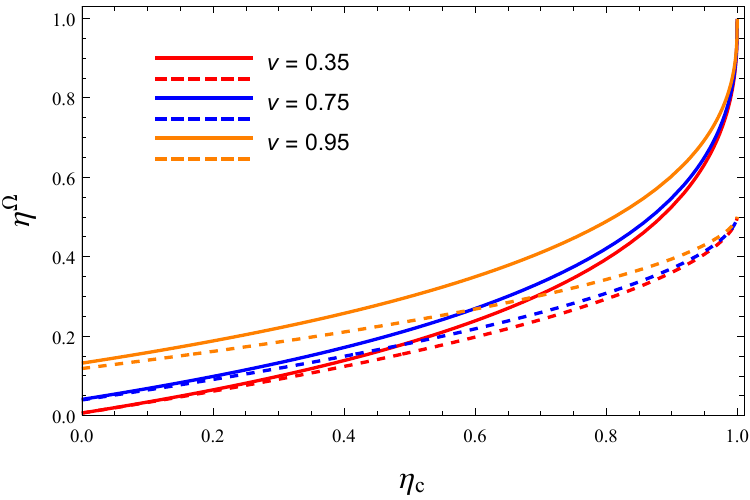}
		\captionsetup{justification=raggedright,singlelinecheck=false}
		\caption{Plot of the efficiency at the maximum Omega function with the Carnot efficiency for different values of oscillator velocity. The solid red, blue, and orange curves represent the efficiency at the maximum $\Omega$ function for the sudden compression case, while the dashed curves in the same colors represent the corresponding efficiency for the sudden expansion case for $v = \{0.35, 0.75, 0.95\}$.}
		\label{fig.2}
	\end{figure}
	The efficiency corresponding to the maximum $\Omega$ function is obtained upon substitution of Eq. (\ref{eq:15}) into Eq. (\ref{eq:11}), and is given by:
	\begin{equation}
		\eta_{SC}^\Omega = \dfrac{(1-J)\left[(1-\eta_{c})(1+J)f_v-2J^2\right]}{(1-\eta_{c})(1+J^2)f_v-2J^2} ,\label{eq:16}
	\end{equation}
	where $ J = z^{*(\Omega)}_{SC}.$ We plot this efficiency expression (Eq. (\ref{eq:16})) in Fig. \ref{fig.2} with the solid red, blue, and orange curves for $v$ = 0.35, 0.75, and 0.95, respectively. From this figure, we observe that the efficiency approaches unity in the sudden compression case and increases with increasing oscillator velocity $v$. Now, we derive the maximum work efficiency by optimizing Eq. (\ref{eq:10}) with respect to $z$, which gives $z^*= (\tau f_v)^{1/3}$. As a result, by using Eq. (\ref{eq:11}), the efficiency at the maximum work is found to be:
	\begin{equation}
		\eta_{SC}^{MW} = \dfrac{K+3L+M\left[2(1-\eta_{c})^5\right]^{1/3}}{K+L-M\left[2(1-\eta_{c})^5\right]^{1/3}}, \label{eq:17}
	\end{equation}
    where $K = 2^{7/3} v (1-\eta_{c})^{2/3}$, $ L = 2(\eta_{c}-1)(v^2 M)^{1/3}$, and $M = x _a \sqrt{1-v^2}$.

	\subsection{Sudden expansion stroke} \label{sb2}
	
	As this scenario complements the previous one, we proceed similarly as before. The roles of the two strokes are now interchanged: the expansion stroke ($C$ $\to$ $D$) is sudden,  with $\lambda_{CD}$ = $(\omega^2_{c}+\omega^2_{h})/{2 \omega_{c} \omega_{h}}$, while the compression stroke ($A$ $\to$ $B$) is adiabatic, with $\lambda_{AB}$ = 1. Within the high-temperature approximation, the expressions for the absorbed heat and extracted work are, respectively, given by:
	\begin{equation}
		Q_{h}^{SE} = \dfrac{1}{z \beta_{h}}(z-\tau f_v),\label{eq:18}
	\end{equation}
	\begin{equation}
		W_{SE}^{HT} = \dfrac{(1-z)}{2z\beta_{h}}\left[z(1+z)-2\tau f_v\right].\label{eq:19}
	\end{equation}
	The corresponding efficiency is given by
	\begin{equation}
		\eta^{HT}_{SE} = \dfrac{(1-z)\left[z(1+z)-2 \tau f_v\right]}{2\left[z-\tau f_v\right]}.\label{eq:20}
	\end{equation}
	Now, to obtain the expression for the maximum efficiency, we optimize Eq. (\ref{eq:20}) with respect to $z$, which gives the following cubic equation:
	\begin{equation}
	4 v^2 z^3 - 3 \tau v x_a z^2 \sqrt{1-v^2} - \tau x_a \left[v\sqrt{1-v^2} + \tau x_a (v^2 - 1) \right] = 0. \label{eq:21}
	\end{equation}
	The above equation is again resolved through the concept of casus irreducibilis as \cite{singh2024asymmetric}
	\begin{equation}
		z^*_{SE} = \dfrac{\tau x_a \sqrt{1-v^2}}{4v}\Biggl\{1+2\cos\left[\dfrac{1}{3}\cos^{-1}\left(\dfrac{\tau^2 x_a^2 (1-v^2)+8 v (v-\tau x_a \sqrt{1-v^2})}{\tau^2 x_a^2 (1-v^2)}\right)\right]  \Biggr\}. \label{eq:22}
	\end{equation}
	\begin{figure}
		\centering
		\includegraphics[width=8.8cm]{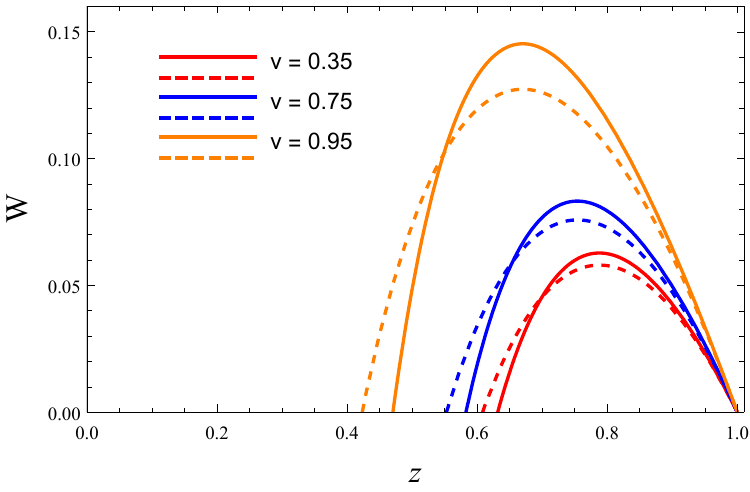}
		\captionsetup{justification=raggedright,singlelinecheck=false}
		\caption{Variation of work output with the ratio of frequencies $z$ for different values of oscillator velocity $v$ and a fixed value of $\tau$ = 0.5. The solid curves correspond to the sudden compression case, while the dashed curves are for the sudden expansion case.}
		\label{fig.3}
	\end{figure}
	Now, we obtain the expression for the maximum efficiency ($\eta_{SE}^{max}$) by using Eq. (\ref{eq:22}) in Eq. (\ref{eq:20}) as:
	\begin{equation}
		\eta^{max}_{SE} = \dfrac{(1-N)\left[N(1+N)-2(1-\eta_{c})f_v\right]}{2\left[N-(1-\eta_{c})f_v\right]} ,\label{eq:23}
	\end{equation}
	where $N = \dfrac{E}{4v}\Biggl\{1+2 \cos \left[\dfrac{1}{3} \cos^{-1} \left(\dfrac{E^2+8v(v-E)}{E^2}\right)\right]\Biggr\}$ and $E = (1-\eta_{c}) x_a \sqrt{1-v^2}$. Next, we use Eqs. (\ref{eq:18}), (\ref{eq:19}), (\ref{eq:23}), and Eq. (\ref{eq:8}) to obtain the expression for the $\Omega$ function corresponding to the sudden expansion stroke. Optimizing the resulting expression with respect to $z$ yields the following optimal value of $z$:
	
	\begin{equation}
		z^{*(\Omega)}_{SE} = \dfrac{\left[\tau x_a \sqrt{1-v^2}
			\left(
			\begin{aligned}
				&x_a \left[3 x_a\tau^2 (v^2-1)\left(2\cos\left(\dfrac{2}{3}P\right)+3\right)
				-32 \tau v \sqrt{1-v^2}\right] \\
				&+4\Bigl[\tau x_a \left(3\tau x_a (v^2-1)+8v\sqrt{1-v^2}\right)
				-24v^2\Bigr]\cos\left(\dfrac{1}{3}P\right)
			\end{aligned}
			\right)\right]^{1/3}}{4 v \left[2\left(1-2\cos \left(\dfrac{1}{3}P\right)\right)\right]^{1/3}},
		\label{eq:24}
	\end{equation}
where $P = \cos^{-1}\left(1-\dfrac{8v\left(v-\tau x_a \sqrt{1-v^2}\right)}{\tau^2 x_a^2(v^2-1)}\right)$. Substituting Eq. (\ref{eq:24}) into Eq. (\ref{eq:20}) yields the efficiency at the optimal $\Omega$ function for the sudden expansion case as follows:
\begin{equation}
	\eta_{SE}^{\Omega} = \dfrac{(1-R)\left[2(1-\eta_{c})f_v - R(R+1)\right]}{2\left[(1-\eta_{c})f_v - R \right]} ,\label{eq:25}
\end{equation}
where $R = z^{*(\Omega)}_{SE} $. Upon optimizing the work output (Eq. (\ref{eq:19})) with respect to $z$ and then using Eq. (\ref{eq:20}), we obtain the efficiency at the maximum work as:
	\begin{equation}
		\eta_{SE}^{MW} = \dfrac{X + 3Y + x_a \sqrt{1-v^2} \left[4(1-\eta_{c})^4\right]^{1/3}}{2(X + Y)}, \label{eq:26}
	\end{equation}
	where $ X = v\left[4(1-\eta_{c})\right]^{1/3} $ and $ Y = (\eta_{c}-1)\left[v x_a^2 (1-v^2)\right]^{1/3} $.

To analyze the engine performance in the sudden expansion case, we plot the efficiency at the maximum $\Omega$ function (Eq. (\ref{eq:25})) using dashed red, blue, and orange for $v$ = 0.35, 0.75, and 0.95, respectively, in Fig. \ref{fig.2}. We observe that in this case, the efficiency attains a maximum value of only one-half. This efficiency behavior arises from the abrupt variation in the oscillator frequency during a sudden stroke, which drives the system out of its instantaneous eigenstate and induces non-adiabatic transitions between the energy levels. These transitions take the system into a non-equilibrium state and generate off-diagonal elements in the density matrix, signaling the emergence of quantum coherence \cite{latune2021roles, PhysRevA.99.062103}. The energy required to create these coherences is stored as parasitic energy and eventually transferred to the thermal reservoirs. This mechanism, referred to as quantum friction, acts as an inherent source of irreversibility and leads to a degradation of the efficiency \cite{rezek2010reflections, plastina2014irreversible, ccakmak2017irreversible, turkpencce2019coupled, insinga2020quantum}.
	
To compare the work output between the sudden compression and sudden expansion cases, we determine the intersection of the corresponding work curves by imposing $W_{SC}^{HT}$ = $W_{SE}^{HT}$ (from Eqs. (\ref{eq:10}) and (\ref{eq:19})) which yields $z = \sqrt{\tau f_v}$. Further, the lower bound of engine operation for the sudden expansion stroke is found by solving $W_{SE}^{HT} = \dfrac{(1-z)}{2z\beta_{h}}\left[z(1+z)-2\tau f_v\right] = 0$, which gives $z = \dfrac{1}{2}\left[\sqrt{1+8 \tau f_v}-1\right] $. The resulting dominance regions, visible in Fig. \ref{fig.3}, are: sudden expansion dominates for $\dfrac{1}{2}\left[\sqrt{1+8 \tau f_v}-1\right]/2 < z < \sqrt{\tau f_v}$, and sudden compression dominates for $ \sqrt{\tau f_v} < z < 1 $. Intersection points of the curves vary with the velocity $ v $ of the oscillator. 
\begin{figure}
	\centering
	\includegraphics[width=8.8cm]{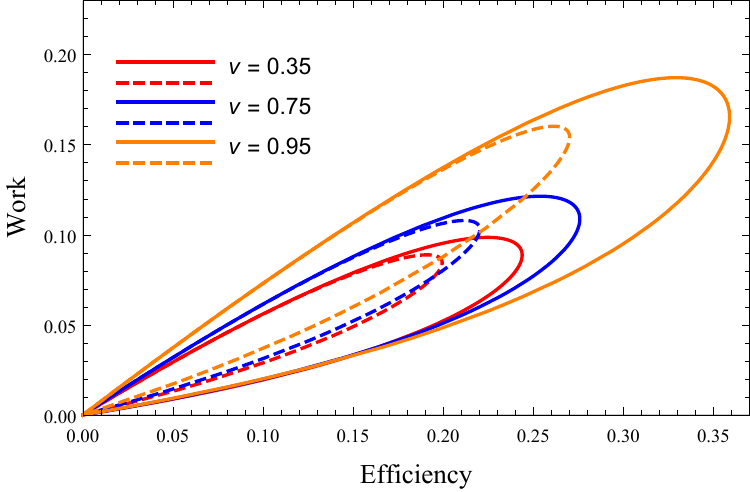}
	\captionsetup{justification=raggedright,singlelinecheck=false}
	\caption{Parametric plots of work versus efficiency for the sudden compression (solid curves) and sudden expansion (dashed curves) cases at a fixed value of $\tau$ = 0.4 and for different $v$ values. The sudden compression results are obtained using Eqs. (\ref{eq:10}) and (\ref{eq:11}), while the sudden expansion results follow from Eqs. (\ref{eq:19}) and (\ref{eq:20}).}
	\label{fig.4}
\end{figure}
	
	In Fig. \ref{fig.4}, we analyze the optimal operational regimes of the sudden compression and the sudden expansion strokes by plotting the efficiency–work curves for $v = 0.35, 0.75,$ and $0.95$ at $\tau = 0.4$. The solid curves correspond to sudden compression [obtained by using Eqs. (\ref{eq:10}) and (\ref{eq:11})], while the dashed ones represent sudden expansion [obtained by using Eqs. (\ref{eq:19}) and (\ref{eq:20})] for different values of oscillator velocity. In both configurations, the loop-like trajectories reflect internal friction arising from the sudden change in frequency. The optimal operating points are identified along the negatively sloped portion of each loop. Notably, increasing the oscillator velocity enlarges the loop area, indicating a simultaneous increase in work output and efficiency. Moreover, for any fixed velocity, the sudden compression protocol consistently surpasses the sudden expansion protocol, implying reduced quantum friction. These observations highlight the pivotal role of asymmetric driving in optimizing the performance of a relativistic quantum Otto cycle.

\section{Complete Phase diagram of Asymmetric Relativistic Quantum Otto Cycle}	\label{sc}

Since frictional effects result in additional operational modes beyond the ideal Otto cycle, we investigate the entire phase diagram of the asymmetric relativistic quantum Otto cycle, which includes four distinct modes, namely the engine, refrigerator, thermal accelerator, and heater \cite{singh2024asymmetric, monika2025asymmetric}. The latter two modes are absent in the quasistatic, frictionless limit and appear only when friction is introduced via finite-time processes. The heat expressions for sudden compression and sudden expansion scenarios are,
	\begin{equation}
		Q^{SC}_{h} = \dfrac{1}{2 z^2\beta_{h}}\left[2z^2 - \tau (z^2+1) f_v\right], \quad Q^{SC}_{c} = \dfrac{1}{\beta_{h}}\left[\tau f_v-z\right].
	\end{equation}
		\begin{equation}
		Q^{SE}_{h} = \dfrac{1}{z \beta_{h}}(z-\tau f_v), \quad Q^{SE}_{c} = \dfrac{1}{\beta_{h}}\left[\tau f_v-\dfrac{1}{2}(1+z^{2})\right].
	\end{equation}

	\begin{table}[t]
		\centering
		\captionsetup{justification=raggedright,singlelinecheck=false}
		\caption{Operational conditions and regimes of the engine, refrigerator, heater, and thermal accelerator modes of the asymmetric relativistic quantum Otto cycle under sudden compression and sudden expansion conditions.}
		\renewcommand{\arraystretch}{1.3}
		\begin{tabular}{p{2.1cm} p{3.2cm} p{6.7cm} p{5cm}}
			\hline 
			\textbf{Mode} & \textbf{Condition} & \begin{tabular}[c]{@{}c@{}}
				\textbf{Sudden Compression} \\
				\textbf{Stroke}
			\end{tabular} &  \begin{tabular}[c]{@{}c@{}}
				\textbf{Sudden Expansion} \\
				\textbf{Stroke}
			\end{tabular}
			\\[2pt]
			\hline 
			\textbf{Engine} & $W_{ext} \geq 0 $, $ Q_{h} \geq 0 $ and $ Q_{c} \leq 0 $ \qquad & $z \geq$ $\dfrac{1}{4}\left[\tau f_v + \sqrt{\tau f_v(\tau f_v+8)}\right] $
			& $z \geq$ $ \dfrac{1}{2}\left[\sqrt{1+8 \tau f_v}-1\right] $ \\
			
			\textbf{Refrigerator} &  $W_{ext} \leq 0 $, $ Q_{h} \leq 0 $ and $ Q_{c} \geq 0 $ \qquad &  $z \leq \tau f_v $ 
			& $ z^2 \leq 2 \tau f_v - 1 $  \\
			
			\textbf{Heater} &  $W_{ext} \leq 0 $, $ Q_{h} \leq 0 $ and $ Q_{c} \leq 0 $ &  $ \tau^2 f_v^2 \leq z^{2} \leq \dfrac{\tau f_v}{2-\tau f_v} $
			& $ 2 \tau f_v - 1 \leq z^2 \leq \tau^2 f_v^2 $ \\
			\textbf{Thermal Accelerator} & $W_{ext} \leq 0 $, $ Q_{h} \geq 0 $ and $ Q_{c} \leq 0 $ & $ \sqrt{\dfrac{\tau f_v}{2-\tau f_v}} \leq z \leq \dfrac{1}{4}\left[\tau f_v + \sqrt{\tau f_v(\tau f_v+8)}\right] $
			& $ \tau f_v \leq z \leq \dfrac{1}{2}\left[\sqrt{1+8 \tau f_v}-1\right] $ \\
			\hline
		\end{tabular}
		\label{tab:1}
	\end{table}
	
		\begin{figure}[H]
		\centering
		\begin{tabular}{ccc}
			\includegraphics[width=0.267\textwidth]{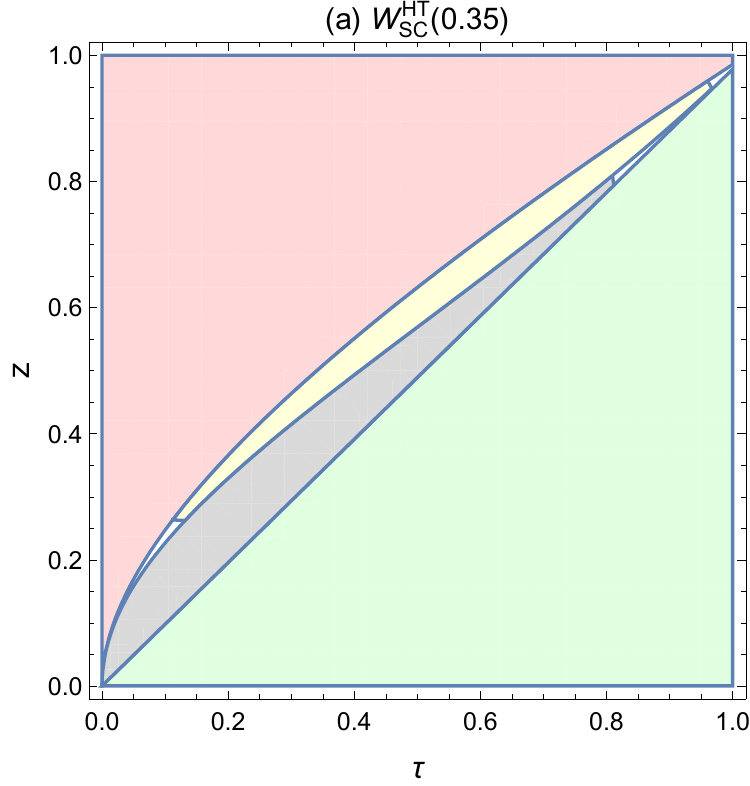} & \quad
			\includegraphics[width=0.267\textwidth]{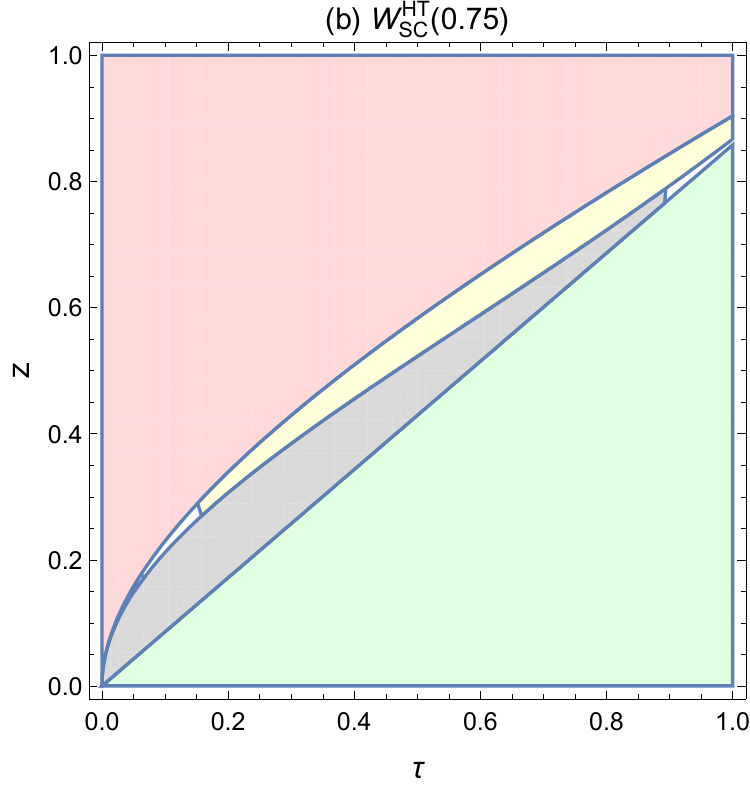} & \quad
			\includegraphics[width=0.36\textwidth,height=4.9cm]{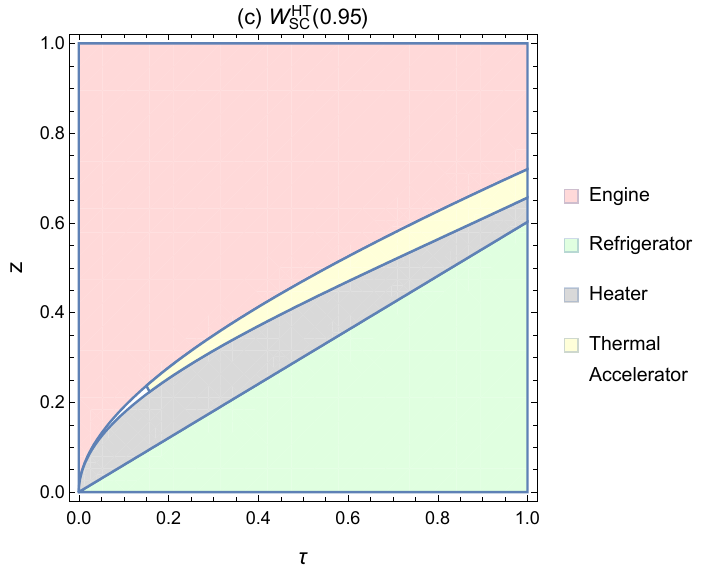}
		\end{tabular}
		\begin{minipage}{\textwidth}
			\caption{Phase diagram of an asymmetric relativistic Otto cycle plotted as a function of $z = \omega_{c}$/$\omega_{h}$ and $\tau$ = $\beta_{h}$/$\beta_{c}$ for sudden compression stroke at the oscillator velocities $v$ = 0.35, $v$ = 0.75, and $v$ = 0.95.}
			\label{fig.5}
		\end{minipage}
	\end{figure}
	\begin{figure}[H]
		\centering
		\begin{tabular}{ccc}
			\includegraphics[width=0.267\textwidth]{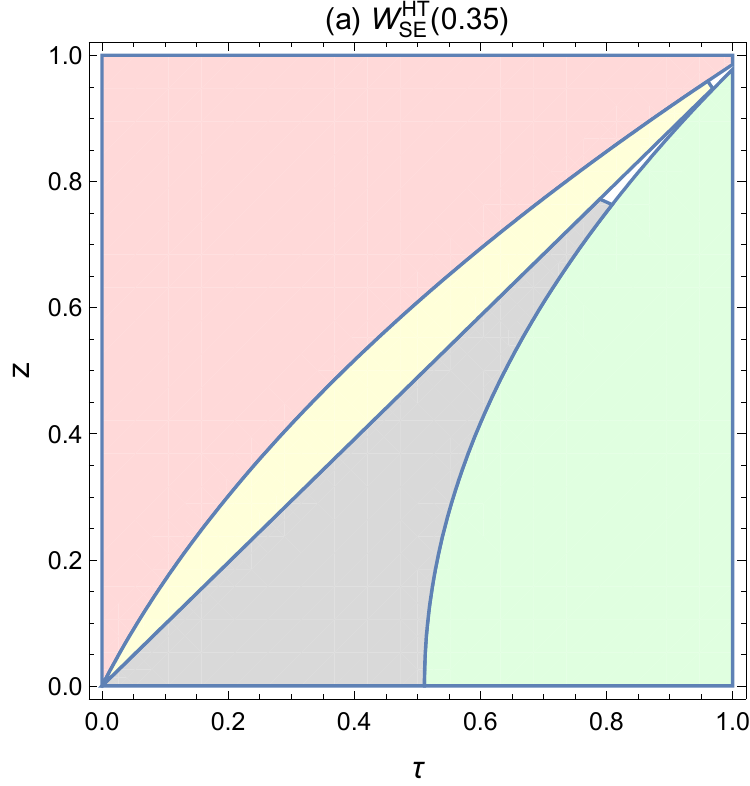} & \quad
			\includegraphics[width=0.267\textwidth]{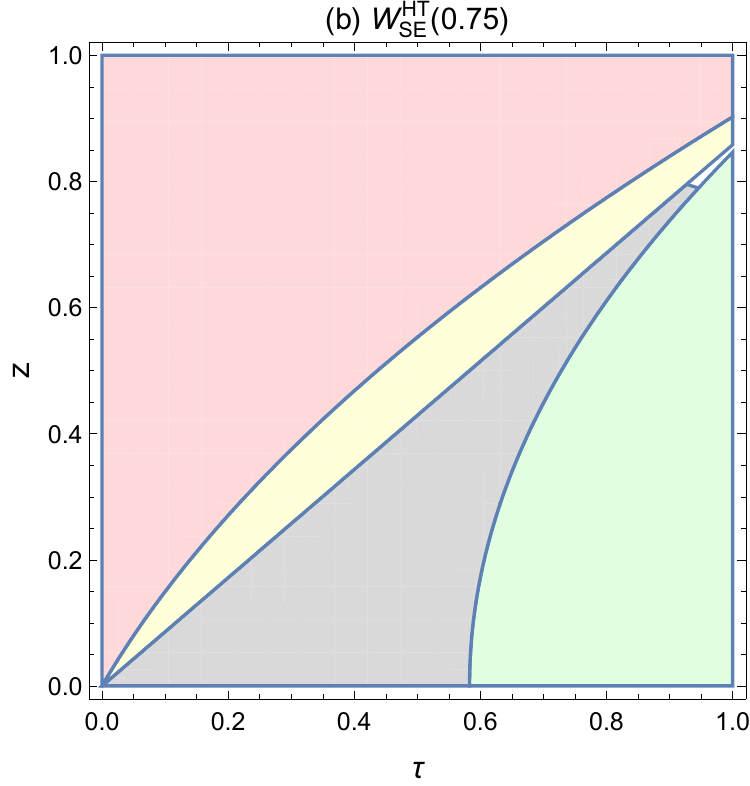} & \quad
			\includegraphics[width=0.36\textwidth,height=4.9cm]{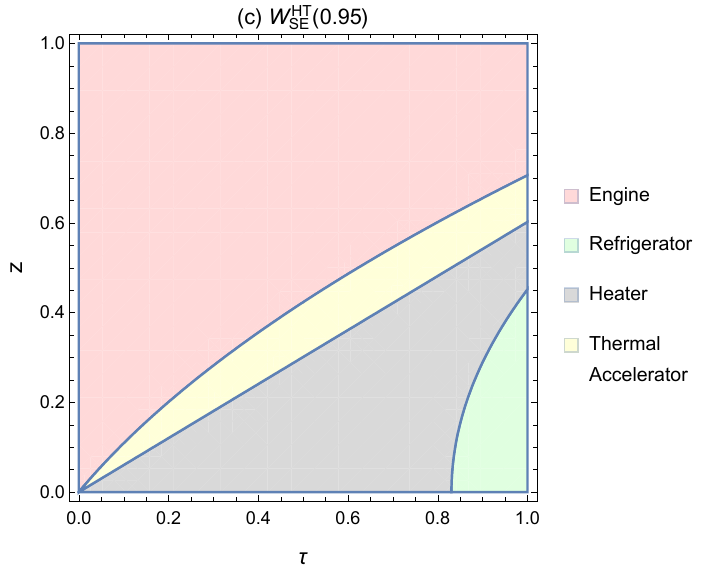}
		\end{tabular}
		\begin{minipage}{\textwidth}
			\caption{Phase diagram of asymmetric relativistic Otto cycle with sudden expansion stroke, plotted against $z = \omega_{c}$/$\omega_{h}$ and $\tau$ = $\beta_{h}$/$\beta_{c}$ and for different values of oscillator velocity, i.e., $v$ = 0.35, 0.75, and 0.95.}
			\label{fig.6}
		\end{minipage}
	\end{figure}
The above equations, together with Eqs. (\ref{eq:10}) and (\ref{eq:19}), are used to identify the various operational modes of the thermal devices in the high-temperature regime. We systematically compile the corresponding results and operating regimes for both the sudden compression and sudden expansion strokes in Table \ref{tab:1}. The phase diagrams in Figs. \ref{fig.5} and \ref{fig.6} reveal a clear trend for both the sudden compression and sudden expansion cases: the engine operational regime expands with increasing oscillator velocity, while the refrigeration regime correspondingly contracts. The heater and accelerator modes are comparatively less sensitive to the oscillator velocity for the parameter values considered here. Additionally, the operating regimes of all the modes depend only on the reservoir temperatures and oscillator velocity.
	
\section{Conclusion} \label{sd}
	
In this work, we examined the performance of a relativistic quantum Otto engine using a harmonic oscillator as the working material. We introduced asymmetry between the two work strokes of the Otto cycle and examined two cases: sudden compression and sudden expansion. First, we found the maximum efficiency for the sudden-compression and sudden-expansion cases. Then, we conducted an optimization study within the framework of the $\Omega$ function and also determined the maximum work efficiency for both the sudden-compression and sudden-expansion scenarios. Our findings revealed that the efficiency at the optimal value of the $\Omega$ function for the sudden compression stroke approaches unity. In contrast, it attains only one-half in the sudden expansion case for all values of velocity. Further, we found a particular range where the engine in the sudden-expansion configuration performs better than in the sudden-compression configuration, and vice versa. Additionally, we examined the optimal operating point for both scenarios by using parametric efficiency-work plots, whose loop-shaped structure indicated that increasing the oscillator velocity enhances both work output and efficiency. Finally, from the full phase diagrams of the Otto cycle, we observe that as oscillator velocity increases, the engine regime expands at the expense of the refrigerator regime for both configurations. Overall, our findings establish a systematic framework for analyzing asymmetric configurations in relativistic quantum Otto cycles. The results provide deeper insights into the role of non-adiabatic effects in the performance of relativistic quantum thermal devices. In this way, our work offers a foundation for future studies on relativistic quantum thermal machines operating beyond the idealized limits of symmetric models.
	
\section*{Acknowledgment}
	
This work is supported by the Government of India through an Institute fellowship at Dr B R Ambedkar National Institute of Technology, Jalandhar.  
	
	\appendix
	
	\section{Casus irreducibilis}
	
	When solving cubic equations, the case of casus irreducibilis may arise when the discriminant, $D = 18abcd-4b^{3}d+b^{2}c^{2}-4ac^{3}-27a^{2}d^{2}$, is positive \cite{benenti2017fundamental}. Consider the general cubic equation,
	
	\begin{equation}
		ay^{3}+by^{2}+cy+d = 0.
	\end{equation}
	We rewrite this equation in the following form:
	
	\begin{equation}
		y^{3}+Ay^{2}+By+C = 0,
	\end{equation}
	where $A = b/a$, $B = c/a$, and $C = d/a$. We express the solution to the above equation using trigonometric functions as
	
	\begin{equation}
		y = -\dfrac{A}{3}+\dfrac{2}{3}\sqrt{A^{2}-3B}\cos\left[\dfrac{1}{3}\cos^{-1}\left(-\dfrac{2A^{3}-9AB+27C}{2(A^{2}-3B)^{3/2}}\right)\right].
	\end{equation}
	
	\subsection{Sudden Compression Case}
	\label{appendixA}
	
	During the sudden compression, the discriminant of the cubic equation
	\begin{equation}
		z^3 + \dfrac{3 \tau x_a \sqrt{1-v^2}}{\tau x_a \sqrt{1-v^2} - 4v} z + \dfrac{(v^2-1)\tau^2 x_a^2 }{\tau v x_a \sqrt{1-v^2} - 4v^2}  = 0, \label{eq:12}
	\end{equation}
	is positive, with $A = 0$, $B = (3 \tau x_a \sqrt{1-v^2})/(\tau x_a \sqrt{1-v^2} - 4v)$, and $C = \left[(v^2-1)\tau^2 x_a^2\right]/(\tau v x_a \sqrt{1-v^2} - 4v^2)$.
	
	\subsection{Sudden Expansion Case}
	\label{appendixB}
	
	During the sudden expansion stroke, the discriminant D for the equation 
	\begin{equation}
		z^3 - \dfrac{3 \tau x_a \sqrt{1-v^2}}{4 v} z^2 - \dfrac{\tau x_a \left[v\sqrt{1-v^2} + \tau x_a (v^2 - 1) \right]}{4 v^2} = 0,
	\end{equation}
	is positive, with $A = -(3 \tau x_a \sqrt{1-v^2})/4 v $, $B = 0$, and $C = - \left[\tau x_a (v\sqrt{1-v^2} + \tau x_a (v^2 - 1))\right]/4 v^2$.
	
%	\section*{References}

\end{document}